\documentclass[aps,prl,reprint,showpacs,superscriptaddress,groupaddress,showkeys,floatfix]{revtex4-1}
\usepackage{amssymb,amsmath}
\usepackage{bm}%
\usepackage[colorlinks=true,linkcolor=blue]{hyperref}
\usepackage{dcolumn}
\usepackage{graphicx}
\usepackage[usenames,dvipsnames]{color}
\expandafter\ifx\csname package@font\endcsname\relax\else
 \expandafter\expandafter
 \expandafter\usepackage
 \expandafter\expandafter
 \expandafter{\csname package@font\endcsname}%
\fi
\begin{document}

\title{Predator-prey dynamics: Chasing by stochastic resetting}

\author{J. Quetzalc\'oatl Toledo-Mar\'in}%
\email{j.toledo.mx@gmail.com}
\affiliation{Instituto de
F\'{i}sica, Universidad Nacional Aut\'{o}noma de M\'{e}xico,
Apartado Postal 20-364, 01000 M\'{e}xico, Ciudad de M\'{e}xico,
M\'{e}xico}%

\author{Denis Boyer}
\email{boyer@fisica.unam.mx}
\affiliation{Instituto de
F\'{i}sica, Universidad Nacional Aut\'{o}noma de M\'{e}xico,
Apartado Postal 20-364, 01000 M\'{e}xico, Ciudad de M\'{e}xico,
M\'{e}xico}

\author{Francisco J. Sevilla}%
\thanks{FJS}
\email{fjsevilla@fisica.unam.mx}
\affiliation{Instituto de
F\'{i}sica, Universidad Nacional Aut\'{o}noma de M\'{e}xico,
Apartado Postal 20-364, 01000 M\'{e}xico, Ciudad de M\'{e}xico,
M\'{e}xico}%

\date{\today}

\begin{abstract}
We analyze predator-prey dynamics 
in one dimension in which a Brownian predator adopts a chasing strategy that consists in 
stochastically resetting its current 
position to locations previously visited by a diffusive prey. 
We study three different chasing strategies, namely, \emph{active}, \emph{uniform} and  \emph{passive} which lead to different diffusive behaviors of the predator in the absence of capture. When 
capture is considered, regardless of the chasing strategy, the mean first-encounter time is finite and decreases with the resetting rate. This model illustrates how the use of cues significantly  improves the efficiency of random searches. We compare 
numerical simulations with analytical calculations and find excellent agreement. 
\end{abstract}

\maketitle
Stochastic processes subject to resetting exhibit diffusive and first passage properties that markedly differ from ordinary diffusion \cite{evans2011diffusion, evans2011diffusionA, EvansJPhysA2013,pal2017first, pal2016diffusion,ReuveniPRL2016, evans2019stochastic}. When a Brownian particle is occasionally reset to a fixed position in space, the mean first passage time to a given target becomes finite and can be minimized with respect to the resetting rate \cite{evans2011diffusion}. Random searches based on resetting principles are advantageous in many contexts \cite{pal2017first, ReuveniPRL2016}, including situations where many targets or resetting points are distributed in space \cite{evans2011diffusionA}.

The statistics of first encounter times is key to understand reaction kinetics between freely diffusive molecules or prey/predators dynamics \cite{krapivsky1996kinetics,redner1999capture, oshanin2009survival, gabel2012can, redner2014gradual}. In the latter context, prey capture can involve relatively complex decisions by the predator depending on the position of the prey, or vice-versa \cite{SchwarzlJPhysA2016,DasJPhysA2018}. For instance, this is the case when the predator uses information
about positions occupied by a prey, and decides to relocate to regions of space where it is more likely to be found \cite{mercado2018lotka}.
There are other phenomena, such as olfaction in the case of olfaction-driven navigation in animals 
\cite{boie2018information, baker2018algorithms, mercado2018lotka}, backtrack recovery in RNA polymerases \cite{dangkulwanich2013complete, roldan2016stochastic} or the formation of physical contacts 
between distant segments of DNA by means of temporal and spatial motion scales \cite{zhang2016first}, that can be modeled by a searcher influenced by cues whose spatial distribution is time dependent.


%
In this Letter we address a problem of two interacting Brownian particles for which the dynamics of one of them, called the 
\emph{predator}, is 
subordinated to the dynamics of the other, called the \emph{prey}, in one-dimensional space. In the following we will phrase the problem in terms of prey and predator for clarity. The chasing dynamics of the predator consist of frequent relocations to positions previously visited by the prey. In other words, this resetting dynamics 
correspond to a non-Markovian search process in which the predator stochastically visits previous prey positions. 
Prey motion is not influenced by the predator here. We focus on the effects of the predator search strategy on its own diffusion and on the statistics of first encounter times with the prey \cite{CamposPRE2017}. Our results may also be relevant in collective animal movement phenomena \cite{ViswanathanBook2011,MendezBook2014, chou2014first, krapf2019strange}.

For this problem, we show that for a finite reset rate, 
the mean capture time is finite, contrary to the situation when the predator 
simply diffuses without resetting for which, as is well known, the mean capture-time diverges 
\cite{redner2014gradual, chou2014first}. 


We model the prey's dynamics as an overdamped Brownian motion of a free particle diffusing in one dimensional space. The time 
evolution of the prey's position, $y(t)$, is given by the stochastic differential equation 
\begin{equation}\label{eq:LangevinY}
 \frac{d}{dt}y(t)=\xi_{y}(t),
\end{equation}
where $\xi_{y}(t)$ denotes a Gaussian-white noise, with mean $\langle\xi_{y}(t)\rangle=0$, and autocorrelation function 
$\langle\xi_{y}(t)\xi_{y}(s)\rangle=2D_{y}\delta(t-s)$, $D_{y}$ being the prey's 
diffusion coefficient and $\delta(t)$ the Dirac's delta function.

\begin{figure}[h]
\includegraphics[width=\columnwidth, trim = {23pt 28pt 20pt 45pt}, clip=true]{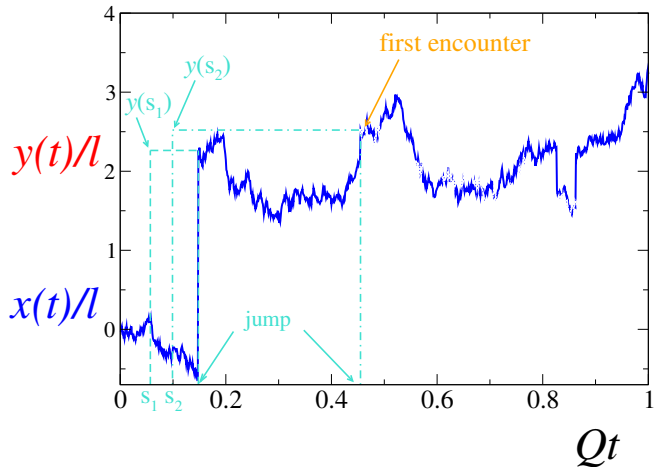}
\caption{Prey and predator dimensionless trajectories, $y(t)/l$ (red), $x(t)/l$ (blue) respectively, for the case when predators and prey diffusion constants are the same 
$D_{x}=D_{y}$. $l$ denotes the length scale $\sqrt{D_{y}/Q}$. The first two jumps of the predator are marked with arrows at $Qt\approx 0.148$ and $Qt\approx0.455$. In the 
first jump the predator choses (from a uniform distribution) to jump to the previously position visited by the prey $y(\text{s}_{1})$ where $\text{s}_{1}$ ($\approx0.057$) is 
chosen from a uniform distribution in the interval $[0,0.148].$ In this example the predator encounters for the first time the prey just right after the second jump (pointed 
with the orange arrow). } \label{fig:BMQ=0.5}
\end{figure}

The predator's dynamics is modeled by the overdamped motion of a Brownian particle that randomly jumps from time to 
time, to a position previously visited by the prey, which makes the subordination process explicit. The time evolution of the predator's position, $x(t)$, is determined by 
the following stochastic differential equation
\begin{equation}\label{eq:LangevinX}
 \frac{d}{dt}x(t)=\xi_{x}(t)[1-\sigma(t)]+\zeta[t,y(\text{s});\text{s}\le t]\sigma(t) \; ,
\end{equation}
that describes the intermittent process of the predator dynamics. We consider the simplest case for the stochastic process $\xi_{x}(t)$, taken as an unbiased 
Gaussian-white noise, i.e. $\langle\xi_{x}(t)\rangle=0$ and $\langle\xi_{x}(t)\xi_{x}(s)\rangle=2D_{x}\delta(t-s)$, where $D_{x}$ denotes the intrinsic diffusion coefficient 
of the predator. $\sigma(t)$ is a dichotomic stochastic process that takes the values 1 at a Poisson rate $Q$.  $\zeta[t,y(\text{s});s\le t]$ denotes the stochastic discontinuous 
process  that describes the predator chasing dynamics, namely, at a constant rate $Q$, the predator jumps from its current position $x(t)$, to a position $y(\text{s})$ 
previously visited by the prey at the random time $\text{s}\le t$ (as depicted in Fig. \ref{fig:BMQ=0.5}), the random variable $\text{s}$ being distributed according to the probability density $\phi(\text{s};t)$. This kernel entails the information that the predator has about past positions of the prey, which we will refer henceforth, as the predator's memory. If the predator has unbiased 
complete memory, any previous time $\text{s}$ is equally probable in the time interval $[0,t]$. Similar memory kernels have been considered in other models, such as the \emph{elephant} random walk 
 \cite{SchutzPRE2004,CressoniPRL2007} or the preferential visit model \cite{boyer2014random}. 

We first focus on the predator dynamics induced by the chasing when no capture of the prey is considered. In such a case, the stochastic processes defined by 
Eqs. \eqref{eq:LangevinY} and \eqref{eq:LangevinX} are equivalently formulated in terms of the conditional 
probability density 
functions $P(y,t\vert y_{0})$, $\Pi(x,t\vert x_{0})$.  The prey's diffusion propagator at time $t$, $P(y,t\vert y_{0})$, is 
given by $G_{D_{y}}(y,t\vert y_{0})=\exp \left\{ -(y-y_0)^2/4D_{y} t\right\}/\sqrt{4\pi D_{y}  t}$, which  is the Gaussian distribution, solution of the diffusion equation 
with diffusion coefficient $D_{y}$ 
and the initial condition $G_{D_{y}}(y,t=0\vert y_{0})=\delta(y-y_{0})$. The predator diffusion propagator, $\Pi(x,t\vert x_{0})$, on the other hand, is given by 
the solution of the Fokker-Planck equation 
\begin{multline}\label{Predator:FPE}
 \frac{\partial}{\partial t} \Pi(x,t\vert x_{0}) =D_x \frac{\partial^{2}}{\partial x^2}\Pi(x,t\vert x_{0})  -  Q\,  \Pi(x,t\vert x_{0})\\
 + Q \int_0^t d\text{s}\, \phi(\text{s};t)\, P(x,\text{s} | y_0),
 \end{multline}
with the initial distribution $\Pi(x,t=0\vert x_{0})=\delta(x-x_{0})$. $\phi(\text{s};t)$ gives the probability density of choosing the time instant 
$\text{s}$ in the interval $[0,t]$.  The first term in the right-hand side of Eq. \eqref{Predator:FPE} corresponds to the predator diffusion process, while the 
second  and third terms refer to the resetting process in which the predator jumps to a position previously visited by the prey, where $P(y,t)dy$ gives the
probability of the prey being at $ \lbrace y, y+dy \rbrace$ at time $t$.  
Equation \eqref{Predator:FPE}  is akin to the continuous-space and continuous-time diffusion process 
under resetting with memory studied in Ref. \cite{BoyerJStatMech2017}. In the present study, the 
subordination to the prey's dynamics leads to new qualitative features as is discussed afterwards.

The statistical properties of the predator diffusion process are derived from $\Pi(x,t\vert x_{0})$. The solution of Eq. \eqref{Predator:FPE} for arbitrary resetting 
strategy $\phi(\text{s};t)$  is given by
\begin{multline}\label{PredatorPDF}
 \Pi(x,t\vert x_{0})=e^{-Qt}G_{D_{x}}(x,t\vert x_{0})+Q\int_{0}^{t}d\text{s}\, e^{-Q(t-s)}\\
 \times\int_{0}^{\text{s}}d\text{s}^{\prime}\phi(\text{s}^{\prime};\text{s})\, G_{D_{y}}\left(x,\left.\frac{D_{x}}{D_{y}}(t-\text{s})+\text{s}^{\prime} \right| y_{0}\right),
\end{multline}
As is expected, the Gaussian distribution 
$G_{D_{x}}(x,t\vert x_{0})$ for the distribution of the predator positions is recovered from \eqref{PredatorPDF} by 
setting 
$Q=0$. At finite $Q$ and in the long-time regime, $Qt \gg 1$, Eq. \eqref{PredatorPDF} reads
\begin{equation}\label{PredatorsPDFAsymptotic}
\Pi_{}(x,t)\sim \int_{-\infty}^{\infty}dx^{\prime}L_{l_{x}}(x-x^{\prime})\int_{0}^{t}d\text{s}\, \phi(\text{s};t)G_{D_{y}}\left(x^{\prime},\text{s}| 
y_{0}\right),
\end{equation}
(see SM in Ref. \cite{SupInfo}). 
$L_{l_{x}}(x)$ denotes the Laplace distribution that occurs in the related diffusion 
process of a Brownian particle that stochastically resets its position to the origin \cite{evans2011diffusion}, given by $\exp\{-\vert x\vert/l_{x}\}/2l_{x}$, 
with $l_{x}=\sqrt{D_{x}/Q}$ 
the characteristic distance the predator travels between consecutive resettings. If this is vanishingly small, i.e., when either the resetting rate is large enough or the 
predator diffusion coefficient 
is small enough, $L_{l_x}(x-x^{\prime})$ becomes sharply distributed around $x$, thus leading to $\Pi_{}(x,t)\sim 
\int_{0}^{t}d\text{s}\, \phi(\text{s};t)G_{D_{y}}\left(x,\text{s}| y_{0}\right)$.

From Eq. \eqref{PredatorPDF} the first two moments can be obtained for arbitrary strategy $\phi(s;t)$, these are given 
explicitly by
\begin{subequations}
 \begin{align}
    \langle x(t)\rangle=&x_{0}e^{-Qt}+y_{0}\left(1-e^{-Qt} \right),\label{PredMeanPosition}\\
     \langle x^{2}(t)\rangle=&x_{0}^{2}e^{-Qt}+\left(y_{0}^{2}+\frac{2D_{x}}{Q}\right)\left(1-e^{-Qt}\right)\nonumber\\
    &\qquad\quad+2D_{y}Q\int_{0}^{t}d\text{s}\, e^{-Q(t-\text{s})}\bar{\tau}(\text{s}),\label{PredMSD}
     \end{align}
 \end{subequations}
where $\bar{\tau}(t)$ denotes the mean time of the distribution $\phi(\tau;t)$ given by the expression
\begin{equation}
 \bar{\tau}(t)=\int_{0}^{t}d\tau\, \tau\, \phi(\tau;t).
\end{equation}
From expression \eqref{PredMeanPosition} it can be deduced that: The  average position of the predator is independent on the resetting strategy $\phi(\text{s};t)$ and tends exponentially fast toward the initial position of the prey, $y_0$; 
in the short-time regime it is given by $x_{0}+(y_{0}-x_{0})Q\, t$, i.e., the predator travels on average ballistically with velocity $(y_{0}-x_{0})Q$.

We now specify these results for an illustrative case, namely, with exponential resetting strategies, where the probability density of picking an instant $\text{s}$ in 
$[0,t]$ is given by
\begin{equation}\label{MemoryStrategy}
 \phi(\text{s};t)=\frac{\lambda e^{-\lambda s}}{1-e^{-\lambda t}},
\end{equation}
with $\lambda$ a real parameter in $(-\infty,\infty)$ that marks the range of the memory. For $\lambda<0$, the chasing strategy is denoted as \textit{active}, \textit{i.e.}, it is based on a short-term memory as 
the predator relocates with a large probability to the most recent positions visited by the prey. The $\lambda=0$ case corresponds to a uniform memory, for which any instant $\text{s}$ in the period of time $[0,t]$ is chosen with the same 
probability weight \cite{boyer2014random,SchutzPRE2004} 
The scenario given by  
$\lambda>0$ corresponds to a \textit{passive} chasing strategy, for which the predator relocates preferentially to the initial positions visited by the prey.

\begin{figure}[t]
\includegraphics[width=3.2in]{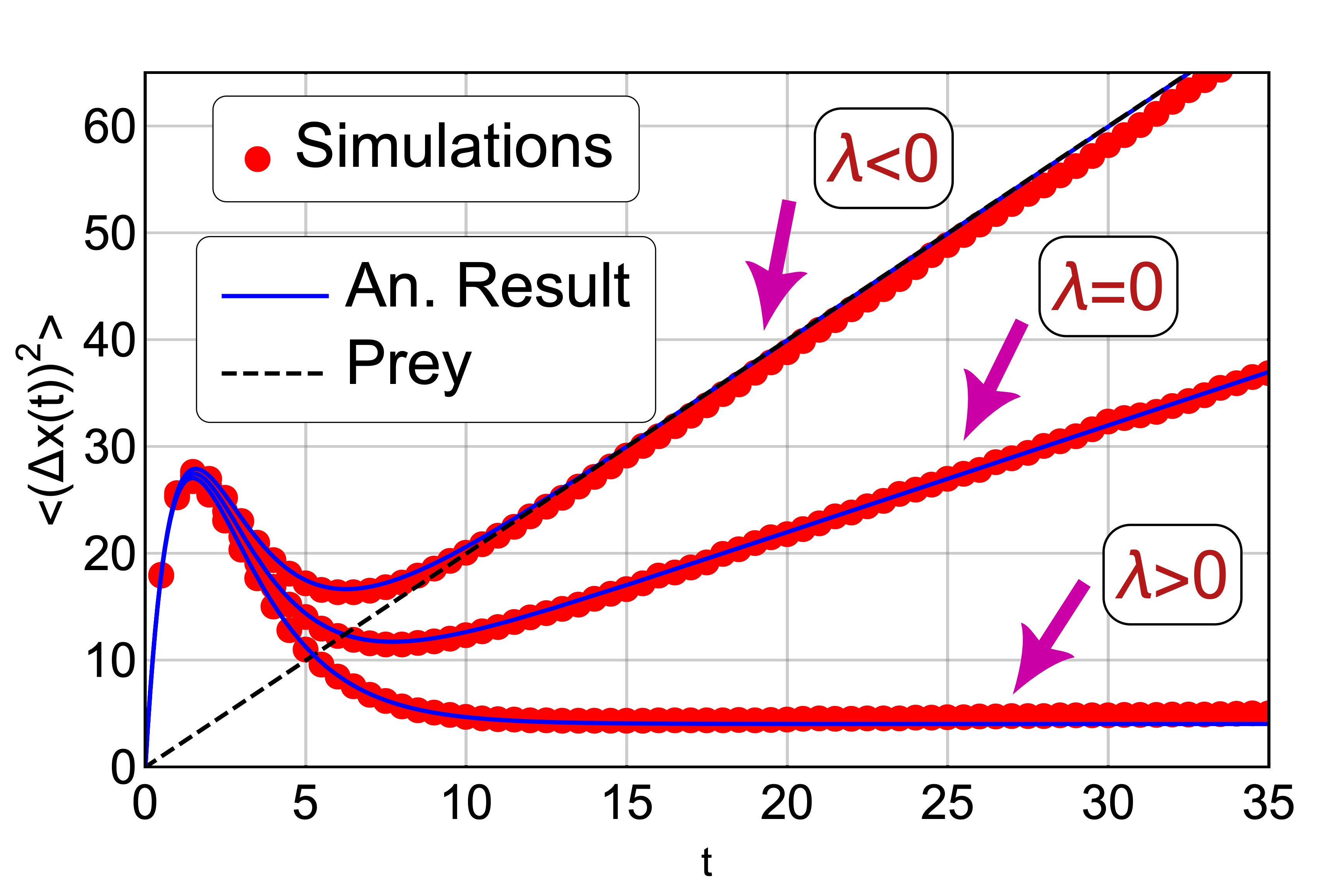}
\caption{\small{ Mean squared displacement of the predator as a function of time with $x_0=0, \; y_0=10, \; D_x=1, \; D_y=1, \; Q=0.5$ and $\langle \left( \Delta x \right)^2 \rangle = \langle x^{2}(t)\rangle - \left(y_{0}^{2}+\frac{2D_{x}}{Q}\right)$. The red 
dots were obtained by simulations, while the blue curves correspond to Eq. (\ref{PredMSD})}. Notice that for large time, i.e., $t \gg 1/Q$, the mean square displacement (MSD) of the 
predator goes as $1/\lambda, \; D_y t$ and $2 D_y t$ when $\lambda >0, \; \lambda =0$ and $\lambda <0$, respectively, as shown in Eq. \eqref{PredMSD}. The MSD 
for the prey (dashed lines) is shown for reference.} \label{fig:MSDUni}
\end{figure}

The predator's 
mean-squared displacement \eqref{PredMSD}, depends on the 
resetting strategy chosen through $\bar{\tau}(t)$. 
For the long-term memory strategy ($\lambda>0$) and $Qt \gg 1$, we have $\bar{\tau}(t) \rightarrow \lambda^{-1}$, and thus the mean-squared displacement saturates $\langle x^{2}(t)\rangle - \left(y_{0}^{2}+\frac{2D_{x}}{Q}\right) \approx 2D_{y}\lambda^{-1}$ (see Fig. \ref{fig:MSDUni}), i.e., the predator gets trapped around the 
prey initial position, similarly to the process with stochastic resetting to the origin \cite{evans2011diffusion}. In the limit 
$\lambda\rightarrow\infty$, $\phi(s;t)\rightarrow\delta(s)$, thus, the predator stochastically resets to the prey's initial position, $y_{0}$, and asymptotically we have 
$\Pi_{}(x)=L_{l_{x}}(x-y_{0})$, which corresponds to the stationary probability distribution found in Ref. \cite{evans2011diffusion}. 

For $\lambda=0$, we have normal diffusion in the large-time limit, since 
$\langle x^{2}(t)\rangle  \sim D_{y}t$, however, the kurtosis of $\Pi_{}(x,t)$ approaches asymptotically to $4$ indicating that it is not Gaussian.
Therefore, this case 
belongs to a class of diffusion processes known as \emph{Brownian yet non-Gaussian diffusion} \cite{SposiniNJPhys2018}, for which the probability distribution is not 
Gaussian in the long-time regime [see Eq. \eqref{PredatorsPDFAsymptotic}]. Remarkably, the predator's diffuses with an effective diffusion coefficient that is half of the 
prey.

For the short-term strategy, $\lambda<0$, the predator jumps to positions recently visited by the prey, 
which at large times yields linear-time dependence $\langle x^{2}(t)\rangle \sim 2D_{y}t$, which indicates that the predator diffuses with the same diffusivity as the prey. In the supplemental material (see Ref. \cite{SupInfo}) we provide the explicit form of the mean squared displacement. 
In Fig. \ref{fig:MSDUni} we compare the time-dependence of the mean-squared displacement obtained from numerical simulations with Eq. \eqref{PredMSD} for which we see an 
excellent agreement.

\begin{figure}[t]
 \includegraphics[width=\columnwidth]{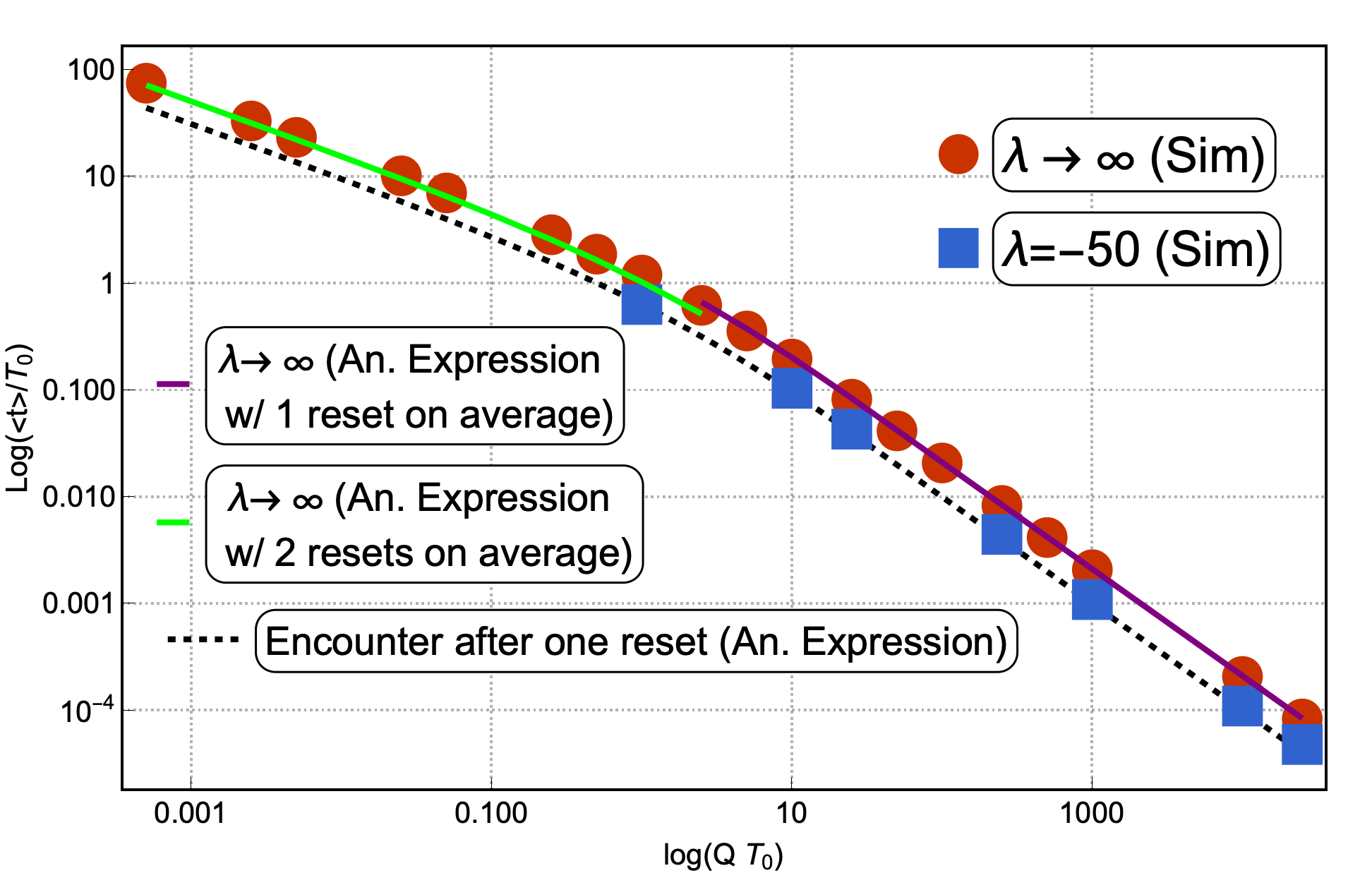}
 \caption{Mean first-encounter time \emph{vs} $Q \mathcal{T}_0$ for different values of $\lambda$ (see legends). 
 The dashed line corresponds to the short-term memory predator resetting strategy $\lambda \rightarrow -\infty$. 
 The blue squares correspond to data obtained from numerical simulations 
 with  $\lambda$ fixed at $-50$ and with time-step size ranging from $10^{-3}$ to $10^{-6}$, depending on the value of $Q$. 
The red circles correspond to numerical simulations 
for which $\lambda \rightarrow \infty$.
The continuous lines correspond to our approximate analytical result (see main text for discussion). Each data point corresponds to $10^7$ simulations. 
} \label{fig:tvsQ}
\end{figure}

We continue our analysis in the scenario for which the predator captures the prey upon first encounter, and we study the statistics of these first encounter times. In the absence of 
the resetting process, the first-encounter time distribution reduces to the L\'evy-Smirnov distribution
$f(t;\mathcal{T}_{0})=\left(\mathcal{T}_{0}/4 \pi t^{3}\right)^{1/2}\exp\left\{-\mathcal{T}_{0}/4t\right\}$,
where $\mathcal{T}_{0}=z_{0}^{2}/(D_{x}+D_{y})$ and $z_0$ is the initial relative distance between the prey and the predator.
In the long-time regime, such distribution is characterized by the 
long tail $f_{0}(t;\mathcal{T}_{0}) \sim t^{-3/2}$, which implies the nonexistence of the mean encounter time \cite{redner2014gradual}. The predator resetting process induces a 
\emph{renewal} of the first-encounter time process, i.e., after the $n$-th resetting event the L\'evy-Smirnov distribution turns into $f(t;\mathcal{T}_{n})$,  where 
$\mathcal{T}_{n}$ is obtained by substituting $z_0$ by $z_n=| x_n-y_n|$, the relative distance between the predator and the prey just right after the stochastic 
relocation of the predator position. Additionally, this \textit{renewal} process frustrates the long tail of the Levy distribution giving way to a finite mean-encounter time. 
We show numerical evidence of this in Fig. \ref{fig:tvsQ} for all values of 
$\lambda$ and finite $Q$, where we have plotted the dimensionless mean encounter time, $\langle t\rangle$. 
We further support this finding with arguments based on approximated analytical calculations.

\begin{figure}[hbtp]
\includegraphics[width=\columnwidth]{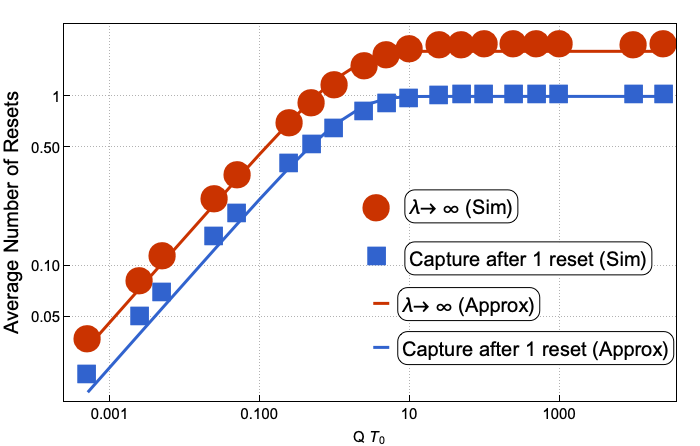}
\caption{\small{Mean number of resets before the first encounter between the prey and predator vs $Q$ for different values of $\lambda$ (see legends). The data points were 
obtained from the simulations while the continuous lines correspond to our analytical results well documented in the SM in Ref. \cite{SupInfo}. There is a threshold for the 
reset rate $Q$ for which above that the mean number of resets before the first encounter saturate. Each point corresponds to $10^7$ simulations.}} \label{fig:resetPlot}
\end{figure}

The mean first-passage time can be computed from the survival probability $\mathcal{S}_Q(z_0,t)$, which can be written as a 
sum, over the number of resets, of the survival probability of a process with exactly $n$ resets. We denote the latter as $\mathcal{S}^{(n)}(z_0,t)$ (see 
the  SM in Ref. \cite{SupInfo} for details on the derivation). For a given sequence of the predator position relocations, $\mathcal{S}^{(n)}(z_0,t)$ may be expressed 
as the convolution of the survival 
probabilities of the diffusive process between two successive reset events, $ \mathcal{S}(z_i,t_i)$, multiplied by the probability that a reset event does not occur. Notice that the survival probability between any two consecutive resets  is simply the survival probability at time $t_i$ of a Brownian particle with initial position $z_i$ and diffusitivity $D=D_x+D_y$, \textit{viz.} $\mathcal{S}(z_i,t_i) = \mathbf{Erf}\left(|z_i|/\sqrt{4D t_i} \right)$, where $\mathbf{Erf}(\bullet)$ is the error function. In Laplace 
domain we have 
 \begin{multline}
\widetilde{ \mathcal{S}}_Q(z_0,\lbrace z_i \rbrace ,u)=\widetilde{\mathcal{S}}(z_0,Q+u) \times\\
\left(1+\sum_{n=1}^{\infty} \prod_{i=1}^n \left[ Q \widetilde{\mathcal{S}}(z_i,Q+u) 
\right] \right) \; . \label{eq:Surv}
 \end{multline}
 In the above expression, we have fixed  $z_i$, which are the relative distance once the $i^{th}$ reset event occurs. The function $\widetilde{\mathcal{S}}(z_0,u)$ is the Laplace transform of the survival probability of the process without 
 reset, namely, 
 \begin{equation}
 \widetilde{\mathcal{S}}(z_0,u)=\frac{1}{u}\left(1-e^{-\sqrt{\frac{u}{D}}\vert z_0 \vert} \right) \; , \label{eq:AsymMFPT}
 \end{equation}
Taking the limit $u \rightarrow 0$ in Eq. \eqref{eq:Surv} yields the mean first-passage time. It is possible to show that when the resetting occurs 
such that $z_i=z_0$ for all $i$, then the summation in Eq. \eqref{eq:Surv} is a geometrical summation and the mean first-passage time obtained coincides with the 
result obtained in Ref. \cite{evans2011diffusion} for the problem of a static prey. In our case, $z_i$ is a random variable and the 
problem becomes analytically intractable, since averages must be performed over the $z_i$'s, of unknown distributions. Nevertheless, we can approximate the sum in Eq. 
\eqref{eq:Surv} by replacing $z_i$ by its typical value $\sqrt{2D_y i/Q}$ (which corresponds to $\lambda \rightarrow \infty$, since the predator relocates to the initial 
position of the prey) and by truncating the summation to $n=1$ and $n=2$ for $Q\mathcal{T}_{0}<1$  and $Q\mathcal{T}_{0}>1$, respectively. This choice is motivated by our simulations in Fig. \ref{fig:resetPlot}, where we have plotted the mean number of resets before the first-encounter \textit{vs} the reset rate obtained from the 
simulations and an analytical approximation documented in the supplemental material \cite{SupInfo}. Figure \ref{fig:tvsQ} displays the resulting mean first encounter times with solid lines, which show excellent agreement with the numerical simulation during seven orders of magnitude in the resetting rate $Q$.

Now notice that for $Q \mathcal{T}_0<1$ the two regimes depend on the predator chasing strategy ($\lambda$) whereas for $Q \mathcal{T}_0 > 1$ this dependence is lost as shown in Fig. \ref{fig:tvsQ}. In addition, $\langle 
t\rangle$ diverges as $(Q\mathcal{T}_{0})^{-1/2}$ for $Q\mathcal{T}_{0}\rightarrow0$, recovering the case at $Q=0$. Around 
$Q\mathcal{T}_{0}\sim 1$, a crossover to the scaling $(Q\mathcal{T}_{0})^{-1}$ is observed  for $Q\mathcal{T}_{0}\gtrsim 1$.  These two scaling regimes are also recovered from the exact analytical expression $\langle t\rangle=\bigl(1-e^{\sqrt{Q\mathcal{T}_{0}}}\bigr)/Q$ for the case $\lambda\rightarrow-\infty$ (dashed line in Fig. \ref{fig:tvsQ}) and 
from Eq. \eqref{eq:Surv} by setting $z_i=0$ for all $i$ and taking the limit $u \rightarrow 0$.
%

Finally, as mentioned, from numerical simulations the average number of resets before the predator-prey encounter, increases from zero with $Q$ and saturates to about $2.2$ (for $\lambda \rightarrow -\infty$) and $1$ (for $\lambda \rightarrow \infty$). We further derived an analytical approximation for the mean number of resets before the predator-prey's encounter (see supplemental material in Ref. \cite{SupInfo} for the analytical derivation). The solid lines in Fig. \ref{fig:resetPlot} corresponds to our analytical expression while the data points corresponds to simulations.

In conclusion, we analyzed the distribution of a Brownian particle that resets to positions previously visited by another Brownian particle. This process depends on a 
memory function that accounts for the available information of the previously visited locations of the prey. We have also studied the first encounter times in this problem. We showed that both particles meet in a finite time, independently of the chasing behavior, and decreases as the 
resetting rate increases. Additionally, the long-time diffusion behavior of the predator is slaved to the diffusion of the prey. 
When only information about the recent locations of the prey is available to 
the predator, the latter tends to mimic the diffusion process of the prey, and ends up diffusing with the prey's diffusion coefficient. If only information of the initial 
positions visited by the prey is available to the predator, the latter becomes trapped around the initial position of the prey. In contrast, if the the whole information is equally available the 
predator ends up diffusing with half the diffusivity of the prey.

\begin{acknowledgments}
F.J.S kindly acknowledges support from grant UNAM-DGAPA-PAPIIT-IN114717. JQTM acknowledges a doctoral fellowship from Consejo Nacional de Ciencia
y Tecnolog\'ia (M\'exico). JQTM gratefully thanks Gerardo G. Naumis, Reinaldo Garcia-Garcia, Mehran Kardar and James Glazier for discussions. 
\end{acknowledgments}

\bibliography{mybib}

\newpage

\appendix

\section{Derivation of Eqs. (5) and (6)}
We explicitly found the solution of the Fokker-Planck equation (4) in the manuscript, namely
\begin{multline}\label{Predator:FPE2}
 \frac{\partial}{\partial t} \Pi(x,t\vert x_{0}) =D_x \frac{\partial^{2}}{\partial x^2}\Pi(x,t\vert x_{0})  -  Q\,  \Pi(x,t\vert x_{0})\\
 + Q \int_0^t d\text{s}\, \phi(\text{s};t)\, P(x,\text{s}).
 \end{multline}
After Fourier transforming this last equation we get
\begin{multline}\label{Predator:FPE-Fourier}
 \frac{\partial}{\partial t} \hat{\Pi}(k,t\vert x_{0}) =-D_x k^{2} \frac{\partial^{2}}{\partial x^2}\hat{\Pi}(k,t\vert x_{0})  -  Q\, \hat{\Pi}(k,t\vert x_{0})\\
 + Q \int_0^t d\text{s}\, \phi(\text{s};t)\, \hat{P}(k,\text{s}),
 \end{multline}
whose solution is given straightforwardly by 
\begin{multline}\label{Predator:FPE-FourierSolution}
 \Pi(x,t\vert x_{0}) =e^{-Qt}\hat{G}_{D_{x}}(k,t)\hat{\Pi}(k,{0})\\
 + Q \int_0^t d\text{s}\, e^{-Q(t-s)}\hat{G}_{D_{x}}\bigl(k,t-s\bigr)\\
 \times\int_{0}^{s}ds^{\prime}\phi(s^{\prime};s)\, \hat{G}_{D_{y}}(k,s^{\prime})e^{-iky_{0}}.
 \end{multline}
where $\hat{G}_{D}(k,t)=\exp\bigl\{-D_{x}k^{2}t\bigr\}$ is the Fourier transform of the Gaussian propagator
\begin{equation}
 G_{D}(u,t\vert u_{0})=\frac{1}{\sqrt{4\pi D  t}} \exp \left\{ -\frac{(u-u_0)^2}{4D t} \right\}
\end{equation}
with diffusion coefficient $D$. By noticing that
\begin{multline}
 \hat{G}_{D_{x}}(k,t)\hat{G}_{D_{y}}(k,s^{\prime})e^{-iky_{0}}=\\
 \hat{G}_{D_{y}}\Biggl(k,\frac{D_{x}}{D_{y}}(t-s)+s^{\prime}\Biggr)e^{-iky_{0}}
\end{multline}
we get the Eq. (5) of the manuscript, namely
\begin{multline}\label{PredatorPDF2}
 \Pi(x,t\vert x_{0})=e^{-Qt}G_{D_{x}}(x,t\vert x_{0})+Q\int_{0}^{t}d\text{s}\, e^{-Q(t-s)}\\
 \times\int_{0}^{\text{s}}d\text{s}^{\prime}\phi(\text{s}^{\prime};\text{s})\, G_{D_{y}}\left(x,\left.\frac{D_{x}}{D_{y}}(t-\text{s})+\text{s}^{\prime} \right| y_{0}\right),
\end{multline}
The analysis in the regime of long times, is easier from Eq. \eqref{Predator:FPE-FourierSolution}. In this regime, the first term in the right-hand side of 
can be neglected, while the second one can be evaluated by use of the \emph{Markovian approximation}, i.e., by taking out the factor that depends on $s^{\prime}$ evaluated at 
$s=t$ multiplied by the integral over $t$ from o to $\infty$ of $Q e^{-Qt}\hat{G}_{D_{x}}\bigl(k,t\bigr)$, i.e.,
\begin{equation}
  \hat{\Pi}_{\infty}(k,t)\sim\frac{Q}{D_{x}k^{2}+Q}\int_{0}^{t}ds\, \phi(s;t)\, \hat{G}_{D_{y}}(k,s)e^{-iky_{0}}.
\end{equation}
Inversion of the Fourier transform leads to the equation (6) of the manuscript, where the Laplace distribution $L_{D_{x}/Q}(x)=\exp\{-\vert x\vert/l\}/2l$ appears as the 
Fourier inverse of $Q/[D_{x}k^{2}+Q]$.

\section{Explicit solutions for particular resetting strategies}
\subsection{Case $\lambda=0$}
We start by formally expressing the problem mathematically, i.e.,
\begin{subequations}\label{FPEQs}
\begin{align}
\frac{\partial}{\partial t}P_0(y,t)& = D_y \frac{\partial^2 }{\partial y^2}P_0(y,t), \label{eq:FPEQsa} \\
\frac{\partial}{\partial t}P(x,t) & = D_x \frac{\partial^2 }{\partial x^2}P(x,t)-QP(x,t) \nonumber \\
&+Q\int_0^t dt' \phi(t,t^{\prime}) P_0(x,t'),  \label{eq:FPEQsb} 
\end{align}
\end{subequations}
 with initial conditions $P_0(y,0)=\delta(y-y_0)$, $P(x,0)=\delta(x-x_0)$ and kernel function $\phi(t,t^{\prime})=1/t$. The solution of Eq. \eqref{eq:FPEQsa} is 
straightforward, namely,
\begin{equation}
P_0(x,t)=\frac{1}{\sqrt{4 D_y \pi t}} \exp \left(-\frac{(y-y_0)^2}{4 D_y t} \right) \; .
\end{equation}
Applying Fourier transform yields
\begin{equation}
P_0(q,t)=  \exp\left( -q^2 D_y  t - \imath q y_0 \right) 
\end{equation}
Now, the general solution to Eq. \eqref{eq:FPEQsb} may be expressed as,
\begin{equation}
P(k,t)=A(t)\exp\left( -(Q+k^2 D_x)t \right) \; . \label{eq:GenSol1}
\end{equation}
Then, when substituting the previous Eq. in the non-homogeneous Focker-Planck Eq. and solving for $A(t)$ yields,
\begin{multline}
A(t)=A(0)\\
+\int_0^t d\tau \exp\left( (Q+k^2 D_x )\tau \right) \frac{Q e^{-\imath k y_0}}{\tau k^2 D_y }\left(1-e^{-k^2 D_y \tau} \right) \; .
\end{multline}
Notice from the initial condition $A(0)=\exp(-\imath k x_0)$. Taking into account the integration constant and the previous Eq., $P(k,t)$ now reads
\begin{multline}
P(k,t)=\exp\left( -(Q+k^2 D_x)t -\imath k x_0 \right) \\
+\int_0^t d\tau \frac{Q e^{-\imath k y_0}}{\tau k^2 D_y }\left(1-e^{-k^2 D_y \tau} \right)\exp\left( -(Q+k^2 
D_x)(t-\tau) 
\right) 
\end{multline}
The first term on the r.h.s. of the last equality is simply the Fourier transform of the solution to the homogeneous Focker-Planck Eq. multiplied by a decaying exponential which makes 
patent the loss in probability in the diffusive regime. Hence, in real space this yield,
\begin{equation}
P_h(x,t)=\frac{e^{-Qt}}{\sqrt{4 \pi D_x t}}\exp\left(-\frac{\left(x-x_0 \right)^2}{4 D_x t } \right) \; .
\end{equation}
The second term is a little bit more complicated to transform into real space. Hence, we provide a detailed way in doing so. Let us first denote this term as $G(k,t)$. Hence,
\begin{multline}
G(k,t)= \int_0^t d\tau \frac{Q e^{Q(\tau-t)}}{\tau} \frac{e^{-\imath k y_0}}{k^2 D_y} \\
 \times \left(e^{-k^2 D_x (t-\tau)}-e^{-k^2 \left(\tau \left(D_y - D_x \right)+D_x t \right)} 
\right) 
\; . \label{eq:G1}
\end{multline}
To obtain the inverse Fourier transform of $G(k,t)$,  we introduce an auxiliary problem. Let us compute the inverse Fourier transform of $\exp \left(-k^2 z\right)/k^2$. Hence, we define a function $F(x)$ such that,
\begin{equation}
F(x)=\frac{1}{2\pi} \int_{-\infty}^{\infty} dk \frac{e^{-k^2 z}}{k^2}e^{\imath k x} \; .
\end{equation}
However, notice that
\begin{multline}
 \frac{d^2F(x)}{dx^2} = -\frac{1}{2\pi} \int_{-\infty}^{\infty} dke^{-k^2 z}e^{\imath k x} 
\end{multline}
The r.h.s in the previous Eq. yields a Gaussian distribution centered at $x=0$ and with root-mean-squared equal to $2z$. Therefore, integrating the above over $x$ and fixing the 
integration constants to the function evaluated in zero yields:
\begin{equation}
\mathcal{F}^{-1} \left[ \frac{e^{- k^2 z}}{k^2} \right](x) \equiv F(x)=-\sqrt{\frac{z}{\pi}}e^{-\frac{x^2}{4z}}- \frac{1}{2}\vert x \vert \mathbf{Erf}\left(\frac{\vert x 
\vert}{2\sqrt{z}}\right) \; .
\end{equation}
We may now go back to our main concern, which is the function $G(k,t)$, and use the previous result by assuming $x \rightarrow x-y_0$. Hence, Eq. \eqref{eq:G1} becomes,
\begin{widetext}
\begin{multline}
G(x,t)= \int_0^t d\tau \frac{Q e^{Q(\tau-t)}}{\tau D_y} \left( -\sqrt{\frac{D_x(t-\tau)}{\pi}}e^{-\frac{(x-y_0)^2}{4D_x (t-\tau)}}- \frac{1}{2}\vert x-y_0 \vert 
\mathbf{Erf}\left(\frac{\vert x-y_0 \vert}{2\sqrt{D_x(t-\tau)}}\right)  \right. \\
 + \left. \sqrt{\frac{\tau (D_y-D_x) +D_x t}{\pi}}e^{-\frac{(x-y_0)^2}{4\left( \tau (D_y-D_x) +D_x t \right)}} + \frac{1}{2}\vert x-y_0 \vert \mathbf{Erf}\left(\frac{\vert 
x-y_0 \vert}{2\sqrt{\tau (D_y-D_x) +D_x t}}\right)  \right) \; .
\end{multline}
By performing further algebra, one is able reduce the previous result to,
\begin{multline}
G(x,t)= \int_0^t d\tau \frac{Q e^{Q(\tau-t)}}{\tau D_y} \left( \sqrt{\frac{\tau (D_y-D_x) +D_x t}{\pi}}e^{-\frac{(x-y_0)^2}{4\left( \tau (D_y-D_x) +D_x t \right)}} 
-\sqrt{\frac{D_x(t-\tau)}{\pi}}e^{-\frac{(x-y_0)^2}{4D_x (t-\tau)}}\right. \\
 - \left.   \frac{1}{2}\vert x-y_0 \vert \mathbf{Erf}\left(\frac{\vert x-y_0 \vert}{2\sqrt{\tau (D_y-D_x) +D_x t}}, \frac{\vert x-y_0 \vert}{2\sqrt{D_x(t-\tau)}} \right)  
\right) \; ,
\end{multline}
and 
\begin{equation}
P(x,t)=\frac{e^{-Qt}}{\sqrt{4 \pi D_x t}}\exp\left(-\frac{\left(x-x_0 \right)^2}{4 D_x t } \right)  + G(x,t) \; . \label{eq:Lambda0}
\end{equation}
This result was compared with simulations and it is shown in Fig. \ref{fig:PDF}.
\end{widetext}

\subsection{Case $\lambda<0$} \label{app:ExpCase}
\begin{figure}[t]
 \includegraphics[width=1.6in]{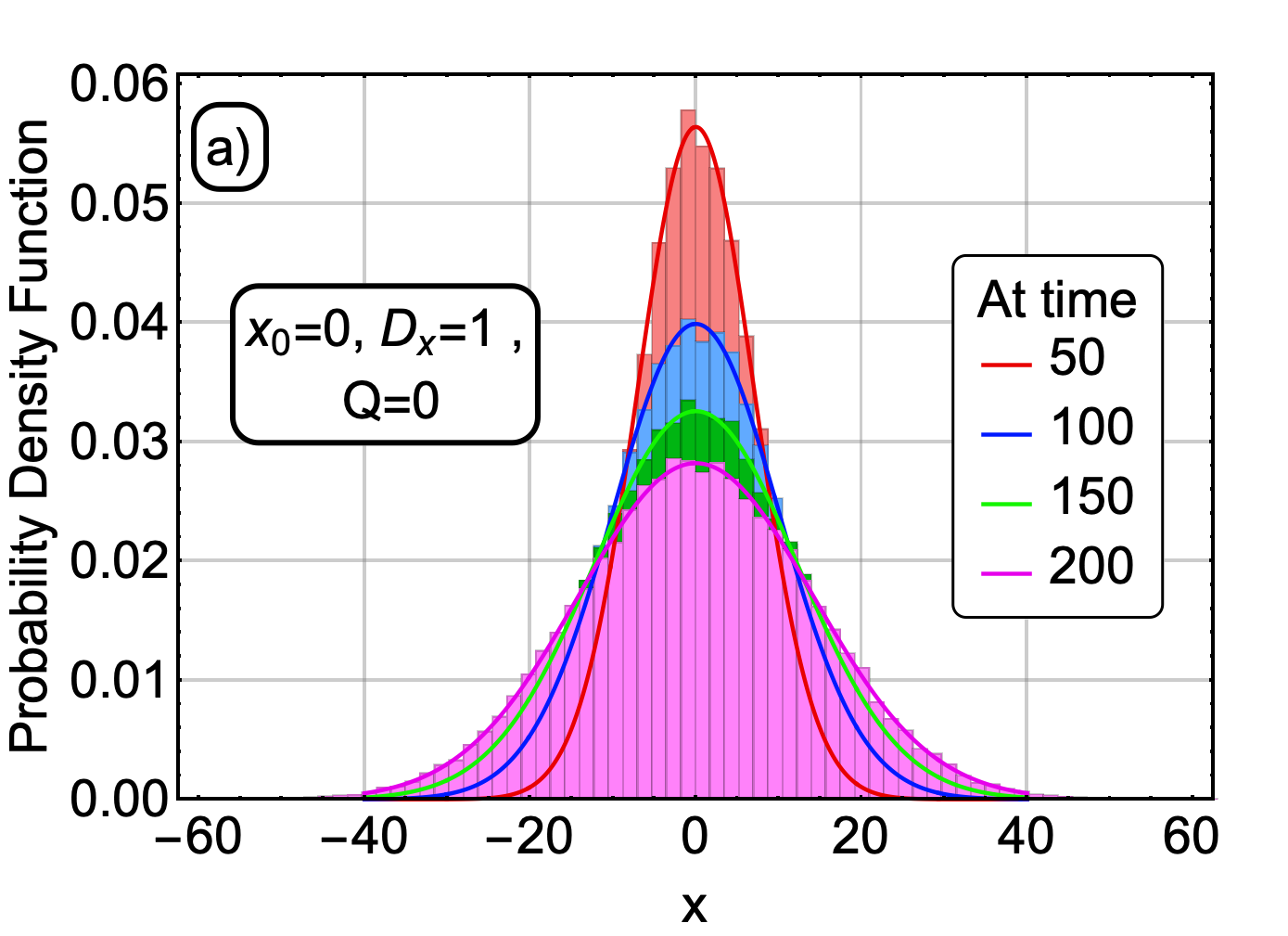}
 \includegraphics[width=1.6in]{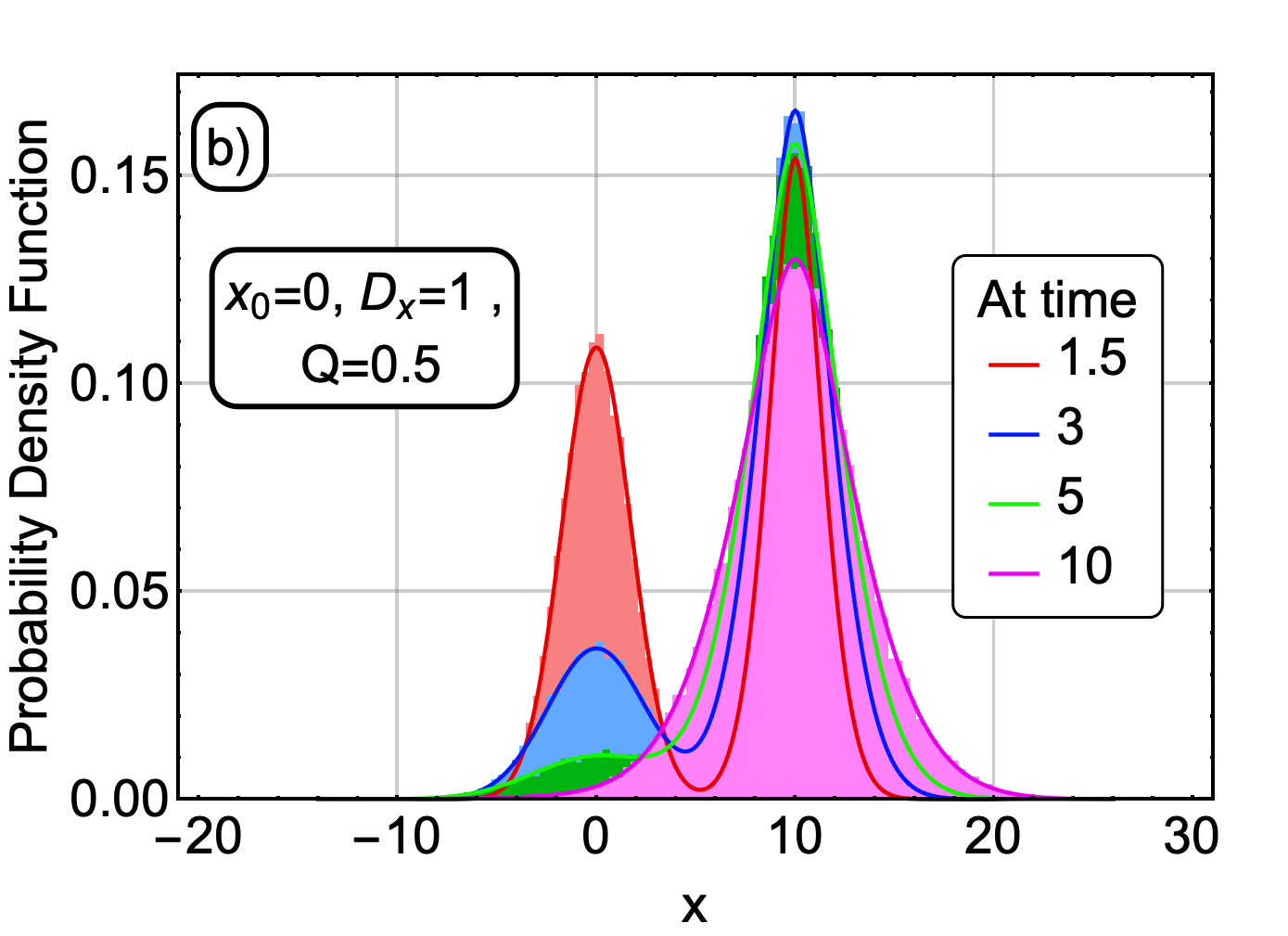}
 \includegraphics[width=1.6in]{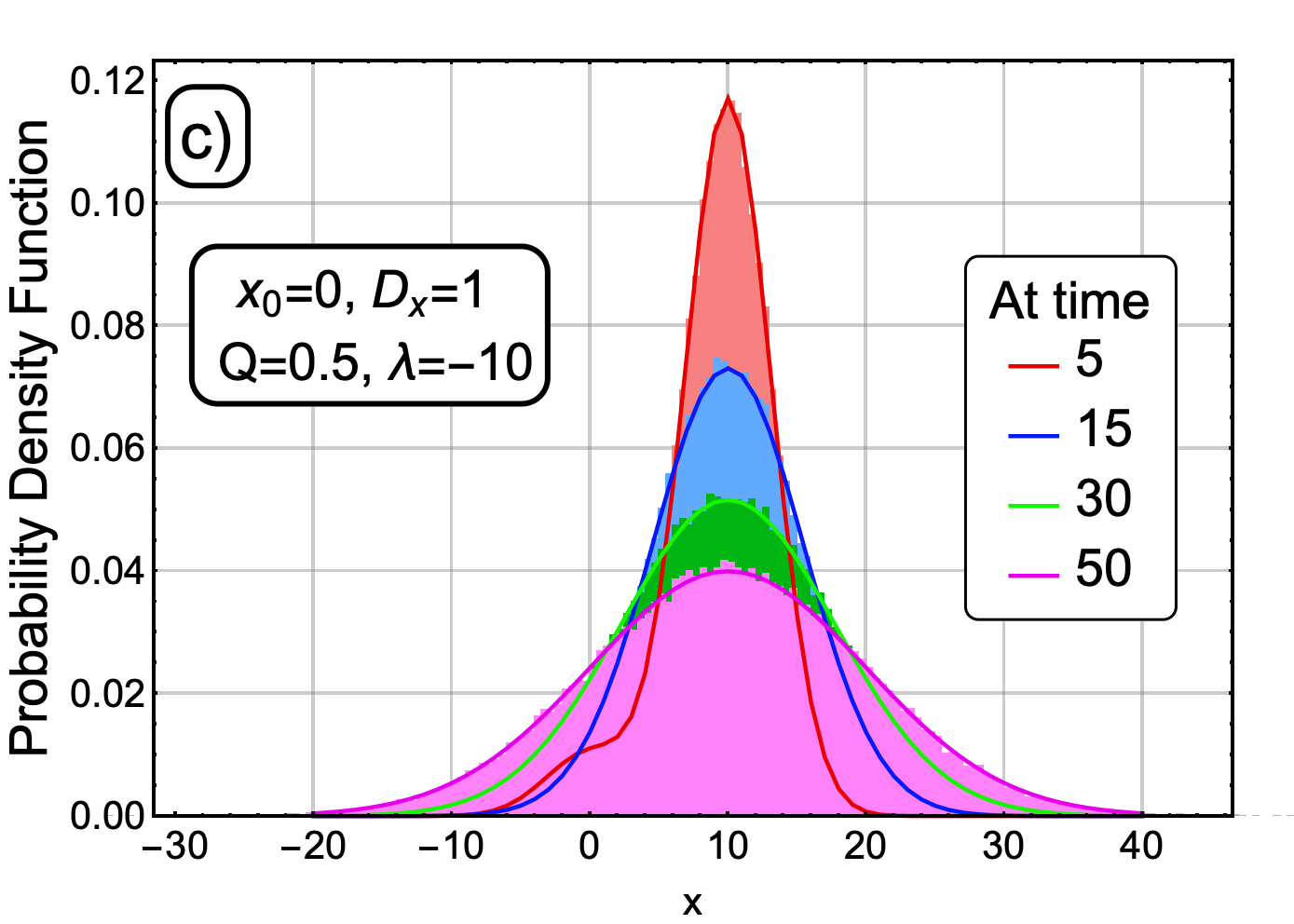}
 \includegraphics[width=1.6in]{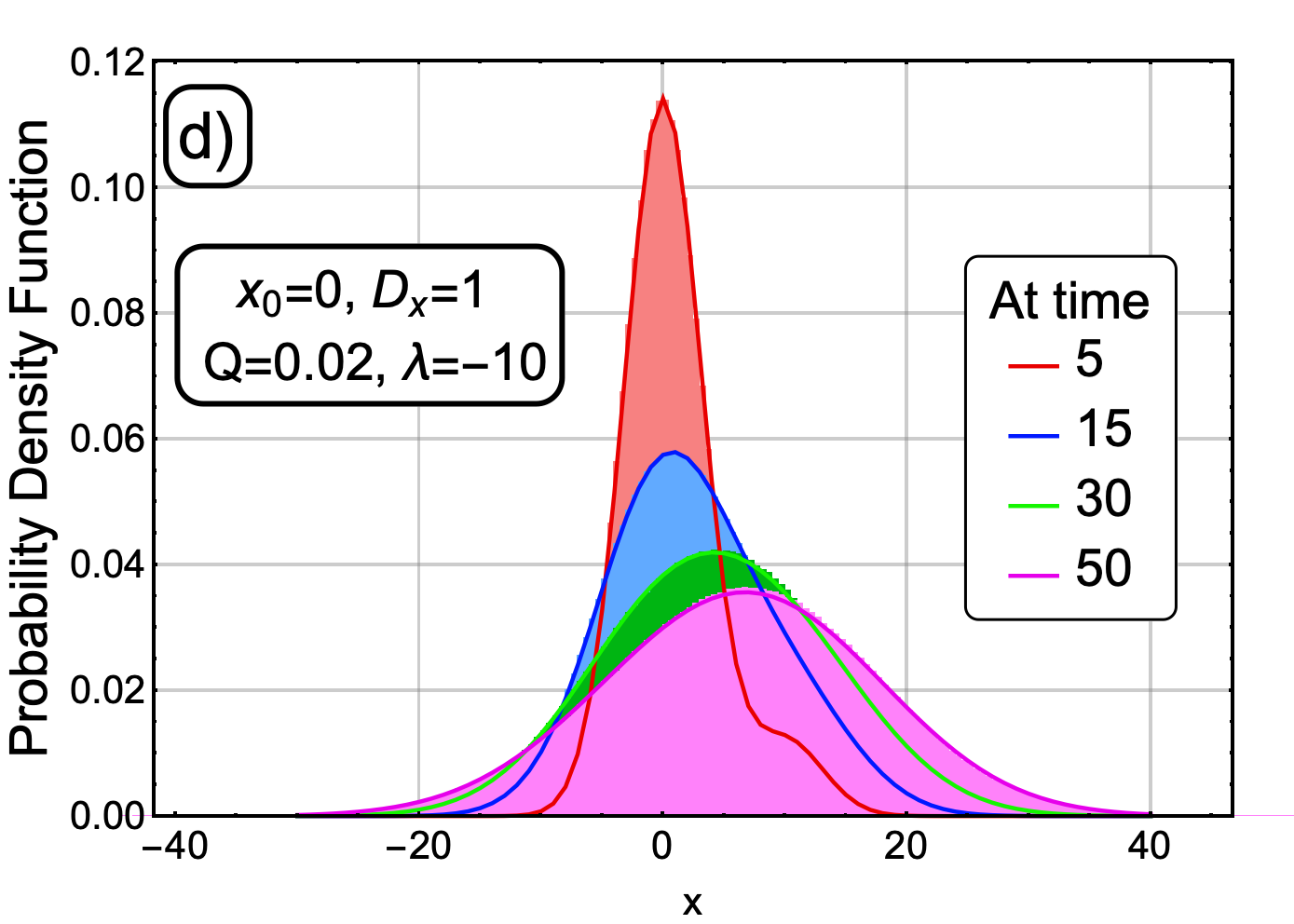}
 \caption{(Color online) Dimensionless probability density function $l_{y}\Pi(x,t\vert x_{0})$ as function of the dimensionless predator position 
 $x/l_{y}$, with $l_y=\sqrt{D_{y}/Q}$. The bars correspond to the histograms obtained from 
 numerical simulations, while 
 solid lines mark the analytical result \eqref{eq:LambdaNeg} and \eqref{eq:Lambda0} (see legends). The simulations were performed fixing the parameters $D_x=D_y, \; x_0=0, \; y_0=10, \; dt=0.001$ and on 
 a 
 sample size $10^4$. Each panel 
 corresponds to a fixed reset rate $Q$ and the different curves within a panel correspond to a different time snapshot of the process, as shown in the 
 legends. Notice as $Q$ increases, the peak shifts towards $y_0$ and becomes sharper.} \label{fig:PDF}
 \end{figure}

Let us first denote $\rho = -\lambda$. Now, we are interested in the explicit solution to Eqs. \eqref{FPEQs} with the same initial conditions as in the previous case but with 
a kernel function,
\begin{equation}
\phi(t,t')=\frac{e^{-\rho(t-t')}}{\Lambda(t)} \; ,
\end{equation}
where
\begin{equation}
\Lambda(t)=\frac{1-e^{-\rho t}}{\rho} \; ,
\end{equation}
 we are interested in the real-space representation of\\
\begin{multline}
P(k,t)= e^{-(Q+k^2 D_x)t-\imath k x_0}  \\
+ \frac{Q \rho e^{-\imath k y_0}}{k^2 D_y - \rho} \int_0^t d\tau \frac{1-e^{-(k^2 D_y - \rho)\tau}}{e^{\rho \tau}-1} 
e^{(Q+k^2 D_x)\tau} e^{-(Q+k^2 D_x)t} \; . \label{eq:Pkt}
\end{multline}
We do this in a similar manner to that in the previous case. For this purpose, let us first derive a formula which will be very useful. Let us suppose we are interested 
in the following integral,
\begin{equation}
F(x; a,b,A, \mathcal{D})=\int_{-\infty}^{\infty} \frac{dk}{2\pi} \frac{e^{-k^2 a}-e^{-k^2 b+A}}{ k^2 - \mathcal{D}} e^{\imath kx} \; .
\end{equation}
Let us define,
\begin{equation}
g(x; a,b,A)= -\int_{-\infty}^{\infty} \frac{dk}{2\pi} \left( e^{-k^2 a}-e^{-k^2 b+A} \right) e^{\imath k x} \; .
\end{equation}
Therefore, the function $F(x)$ satisfies the following differential equation,
\begin{equation}
F''(x)+DF(x)=g(x) \; .
\end{equation}
Now, to solve the previous Eq., one might be tempted in using Fourier Transform method. However, that would take us back to our starting point. Hence, we use the parameter 
variation method \cite{elsgolts1977differential}. Thus, the solution yields
\begin{equation}
F(x; a,b,A, \mathcal{D})=c_1(x) \cos \left(\sqrt{ \mathcal{D}}x \right)+ c_2(x) \sin \left(\sqrt{ \mathcal{D}}x \right) \; .
\end{equation}
where,
\begin{equation}
\begin{cases}
c_2(x; a,b,A, \mathcal{D})=\int_{0}^x dx' \frac{g(x')\cos \left(\sqrt{ \mathcal{D}}x' \right)}{\sqrt{ \mathcal{D}}}  \; , \\
c_1(x; a,b,A, \mathcal{D})= -\int_{-\infty}^x dx' \frac{g(x')\sin \left(\sqrt{ \mathcal{D}}x' \right)}{\sqrt{ \mathcal{D}}}  \; .
\end{cases}
\end{equation}
Notice that we have chosen different integration limits. The reason will become clear later on. Therefore, we have
\begin{multline}
F(x; a,b,A, \mathcal{D})= 
-\int_{-\infty}^x dx' \frac{g(x')\sin \left(\sqrt{ \mathcal{D}}x' \right)}{\sqrt{ \mathcal{D}}} \cos \left(\sqrt{ \mathcal{D}} x 
\right) \\
+ \int_{0}^x dx' \frac{g(x')\cos \left(\sqrt{ \mathcal{D}}x' \right)}{\sqrt{ \mathcal{D}}} \sin \left(\sqrt{ \mathcal{D}} x \right) \; .
\end{multline}
Notice that when $ \mathcal{D}$ and $A$ tend to $0$, the previous Eq. becomes
\begin{multline}
F(x;a,b,A, \mathcal{D})=
 \frac{\sqrt{b}}{\sqrt{\pi}} e^{-\frac{x^2}{4b}} - \frac{\sqrt{a}}{\sqrt{\pi}} e^{-\frac{x^2}{4a}}\\
  - \frac{x}{2} \left(\mathbf{Erf}\left(\frac{x}{\sqrt{4b}}, 
\frac{x}{\sqrt{4a}} \right) \right) \; .
\end{multline}
In agreement with the case in the previous section.
Let us return to to Eq. \eqref{eq:Pkt} and rewrite it as
\begin{widetext}
\begin{multline}
P(k,t)= e^{-(Q+k^2 D_x)t-\imath k x_0} + \int_0^t d\tau  \frac{Q \rho e^{-Q(t-\tau)} }{\left(e^{\rho \tau}-1\right) D_y} e^{-\imath k y_0} F(k; D_x(t-\tau), D_y \tau 
+D_x(t-\tau), \rho \tau ,\rho/D_y ) \; .
\end{multline}

Using the previous results leads to
\begin{eqnarray}
P(x,t)&=&\frac{e^{-Qt}e^{-\frac{(x-x_0)^2}{4D_x t}}}{\sqrt{4\pi D_x t}} + \int_0^t d\tau  \frac{Q \rho e^{-Q(t-\tau)} }{\left(e^{\rho \tau}-1\right) D_y} F(x-y_0; 
D_x(t-\tau), D_y \tau +D_x(t-\tau), \rho \tau ,\rho/D_y )
\end{eqnarray}

Let us now check the normalization condition in the previous Eq. by integrating over the real axis:

\begin{equation}
1=\int_{-\infty}^{\infty}dxP(x,t)=e^{-Qt}+ \int_0^t d\tau  \frac{Q \rho e^{-Q(t-\tau)} }{\left(e^{\rho \tau}-1\right) D_y} \int_{-\infty}^{\infty}dx F(x-y_0; 
D_x(t-\tau), D_y \tau +D_x(t-\tau), \rho \tau ,\rho/D_y ) \label{eq:NormalizationExpTrail}
\end{equation} 
\end{widetext}
Now, the dependence on $x$ in the second term is solely because of $F(x)$. Therefore, let us integrate $F(x)$. Let us denote the terms in $F(x)$ as
\begin{equation}
\begin{cases}
G_s(x)=\int_{-\infty}^x dx' \frac{g(x')\sin \left(\sqrt{ \mathcal{D}}x' \right)}{\sqrt{ \mathcal{D}}} \; , \\
G_c(x)=\int_{0}^x dx' \frac{g(x')\cos \left(\sqrt{ \mathcal{D}}x' \right)}{\sqrt{ \mathcal{D}}} \; ,
\end{cases}
\end{equation}
Thus,
\begin{multline}
\int_{-\infty}^{\infty} dx F(x) = - \int_{-\infty}^{\infty} G_s(x) \cos \left(\sqrt{ \mathcal{D}} x \right) \\
+ \int_{-\infty}^{\infty} G_c(x) \sin \left(\sqrt{ \mathcal{D}} x \right) \; .
\end{multline}
Integrating by parts and taking into account that $dG_s(x)/dx$ is an odd function, yields
\begin{multline}
\int_{-\infty}^{\infty} dx F(x) = \frac{1}{ \mathcal{D}} \int_{-\infty}^{\infty}g(x) \\
 + \left. \frac{G_c(x) \cos \left(\sqrt{ \mathcal{D}} x \right)}{\sqrt{ \mathcal{D}}} \right|_{-\infty}^{\infty} \; .
\end{multline}
It is not difficult to see that the last term is zero, given the way we have chosen the integration limits of $G_c(x)$. Hence,
\begin{eqnarray}
\int_{-\infty}^{\infty} dx F(x) = \frac{1}{ \mathcal{D}} \int_{-\infty}^{\infty}g(x) = \frac{1}{ \mathcal{D}}\left(e^{A}-1 \right) \; .
\end{eqnarray}
Plugging this in Eq. (\ref{eq:NormalizationExpTrail}) with $A=\rho \tau$ and $ \mathcal{D}=\rho/D_y$, fulfills the equality.  
The integrals in the previous results may further be put in terms of Error and Dawson functions. Before carrying on, let us first write some useful identities, namely:
\begin{widetext}
\begin{eqnarray}
\int_0^x dx' \frac{e^{-\frac{x'^2}{4 \alpha}}}{\sqrt{4\pi \alpha}} e^{\pm \imath k x'} &=& \frac{e^{-k^2 \alpha}}{2} \mathbf{Erf}\left(\mp \imath \sqrt{\alpha} k, \frac{x\mp 
\imath 2 \alpha k}{\sqrt{4\alpha}} \right) \nonumber \; , \\
\int_0^x dx' \frac{e^{-\frac{x'^2}{4 \alpha}}}{\sqrt{4\pi \alpha}} \cos k x' &=& \frac{1}{2} \frac{e^{-k^2 \alpha}}{2} \left( \mathbf{Erf}\left( \frac{x- \imath 2 \alpha 
k}{\sqrt{4\alpha}} \right) + \mathbf{Erf}\left(\frac{x + \imath 2 \alpha 
k}{\sqrt{4\alpha}} \right) \right) \nonumber \; , \\
\int_0^x dx' \frac{e^{-\frac{x'^2}{4 \alpha}}}{\sqrt{4\pi \alpha}} \sin k x' &=& \frac{1}{2 \imath} \frac{e^{-k^2 \alpha}}{2} \left( \mathbf{Erf}\left( -\imath \sqrt{\alpha} 
k, \frac{x- \imath 2 \alpha k}{\sqrt{4\alpha}} \right) - \mathbf{Erf}\left( \imath \sqrt{\alpha} k, \frac{x + \imath 2 \alpha k}{\sqrt{4\alpha}} \right) \right) \nonumber \; 
, 
\\
\int_0^{\infty} dx' \frac{e^{-\frac{x'^2}{4 \alpha}}}{\sqrt{4\pi \alpha}} \sin k x' &=& \frac{\mathbf{D}_+\left(k\sqrt{\alpha} \right)}{\sqrt{\pi}} \; .
\end{eqnarray}

Here, $\mathbf{D}_+(x)$ is the Dawson function. Hence, each of the terms in the function $F(x;a,b,A,\mathcal{D})$ may be expressed in terms of the previous Eqs., i.e.,
\begin{multline}
\int_0^{x} dx' \frac{g(x') \cos \left(\sqrt{ \mathcal{D}}x'\right)}{\sqrt{ \mathcal{D}}} = \frac{1}{4\sqrt{ \mathcal{D}}} \left(e^{A-\mathcal{D}b} \left(  \mathbf{Erf}\left( 
\frac{x- \imath 2 b \sqrt{ \mathcal{D}}}{\sqrt{4b}} 
\right) + \mathbf{Erf}\left(\frac{x + \imath 2 b \sqrt{ \mathcal{D}}}{\sqrt{4b}} \right) \right) \right. \\
 \left. - e^{-\mathcal{D}a} \left(  \mathbf{Erf}\left( \frac{x- \imath 2 a \sqrt{ \mathcal{D}}}{\sqrt{4a}} \right) + 
\mathbf{Erf}\left(\frac{x + \imath 2 a \sqrt{ \mathcal{D}}}{\sqrt{4a}} \right) \right)\right)
\end{multline}
\begin{eqnarray}
\int_0^{x} dx' \frac{g(x') \sin \left(\sqrt{ \mathcal{D}}x'\right)}{\sqrt{ \mathcal{D}}} &=& \frac{\imath}{4\sqrt{ \mathcal{D}}} \left(e^{A-\mathcal{D}b} \left( - 
\mathbf{Erf}\left( \frac{x- \imath 2 b 
\sqrt{ \mathcal{D}}}{\sqrt{4b}} \right) + \mathbf{Erf}\left(\frac{x + \imath 2 b \sqrt{ \mathcal{D}}}{\sqrt{4b}} \right)-2 \mathbf{Erf} \left(\imath \sqrt{\mathcal{D}b} 
\right) \right) \right. \nonumber \\
&& \left.- e^{-\mathcal{D}a} \left( - \mathbf{Erf}\left( \frac{x- \imath 2 a \sqrt{ \mathcal{D}}}{\sqrt{4a}} \right) + \mathbf{Erf}\left(\frac{x + \imath 2 a \sqrt{ 
\mathcal{D}}}{\sqrt{4a}} \right) -2 
\mathbf{Erf} \left(\imath \sqrt{Da} \right) \right)\right)
\end{eqnarray}
\begin{equation}
\int_{-\infty}^{0} dx' \frac{g(x') \sin \left(\sqrt{ \mathcal{D}}x'\right)}{\sqrt{ \mathcal{D}}} 
=- \frac{1}{\sqrt{ \mathcal{D} \pi}} \left(e^{A} \mathbf{ \mathcal{D}}_+(\sqrt{Db}) - \mathbf{ \mathcal{D}}_+(\sqrt{Da}) \right) \; .
\end{equation}

Therefore, after some algebra, we obtain
\begin{eqnarray}
F(x;a,b,A, \mathcal{D})&=&  \frac{e^{A-Db}}{\sqrt{4D}} \mathbf{Im}\left[\mathbf{Erf}\left(\frac{x+\imath 2 b \sqrt{ \mathcal{D}}}{\sqrt{4b}} \right) e^{\imath \sqrt{ 
\mathcal{D}}x} \right]  -  
\frac{e^{Da}}{\sqrt{4D}} \mathbf{Im}\left[\mathbf{Erf}\left(\frac{x+\imath 2 a \sqrt{ \mathcal{D}}}{\sqrt{4a}} \right) e^{\imath \sqrt{ \mathcal{D}}x} \right] \; .
\end{eqnarray}
Thus, by putting the parameters $a=D_x(t-\tau), \; b=D_y \tau + D_x(t-\tau),\; A=\rho \tau$ and $\mathcal{D}=\rho/D_y$, we obtain:
\begin{multline}
F(x-y_0;D_x(t-\tau),D_y \tau + D_x(t-\tau),\rho \tau,\rho/D_y) =  \frac{e^{\rho \tau-\rho\left(D_y \tau + D_x(t-\tau) \right)/D_y}}{\sqrt{4\rho/D_y}} \\
\mathbf{Im}\left[\mathbf{Erf}\left(\frac{x-y_0+\imath 2 \left( D_y \tau + D_x(t-\tau) \right) \sqrt{\rho/D_y}}{\sqrt{4\left(D_y \tau + D_x(t-\tau) \right)}} \right) 
e^{\imath \sqrt{\rho}(x-y_0)/\sqrt{D_y}} \right] \nonumber \\
  -  \frac{e^{-\rho D_x(t-\tau)/D_y}}{\sqrt{4\rho/D_y}} \mathbf{Im}\left[\mathbf{Erf}\left(\frac{x-y_0+\imath 2 D_x(t-\tau) \sqrt{\rho/D_y}}{\sqrt{4D_x(t-\tau)}} 
\right) e^{\imath \sqrt{\rho}(x-y_0)/\sqrt{D_y}} \right] \; ,
\end{multline}
and
\begin{eqnarray}
P(x,t)&=&\frac{e^{-Qt}e^{-\frac{(x-x_0)^2}{4D_x t}}}{\sqrt{4\pi D_x t}} + \int_0^t d\tau  \frac{Q \rho e^{-Q(t-\tau)} }{\left(e^{\rho \tau}-1\right) D_y} F(x-y_0; 
D_x(t-\tau), D_y \tau +D_x(t-\tau), \rho \tau ,\rho/D_y ) \; .  \label{eq:LambdaNeg}
\end{eqnarray}
\end{widetext}
This result was compared with simulations and it is shown in Fig. \ref{fig:PDF}.

\subsection{Mean value and mean squared displacement}

Let us derive the mean value and the mean squared displacement (MSD). There are several ways on doing this, yet a simple way consists in multiplying the Fokker-Planck Eq. by $x^n$, integrating it over the Real line and 
then solving the resulting differential equation.
In the case of the mean value, we must solve the following differential Eq.
\begin{equation}
\frac{d \langle x (t) \rangle}{dt}=-Q\langle x(t) \rangle +Q\int_0^t dt' \phi(t,t') y_0 = -Q\langle x(t) \rangle +Q y_0  \; .
\end{equation}
The solution to the previous Eq. is quite simple and independent of the memory kernel, i.e.,
\begin{equation}
\langle x(t) \rangle = x_0 e^{-Qt}+ y_0 \left(1-e^{-Qt} \right) \; .
\end{equation}
For the MSD, denoting $\eta(t)=\langle x^2(t) \rangle$ to simplify notation, following the same procedure we obtain the following differential Eq.
\begin{equation}
\frac{d \langle x^2(t) \rangle}{dt}=-Q\langle x^2(t) \rangle+2D_x + Q \int_0^t dt' \phi(t,t') \left(y_0^2+2D_y t' \right) \; .
\end{equation}
The solution is straightforward, namely,
\begin{multline}
\langle x^2(t) \rangle _{\lambda <0} = x_0^2 e^{-Qt} + y_0^2 \left(1-e^{-Qt} \right) + \frac{2D_x}{Q} \left(1-e^{-Qt} \right)  \\
+ \frac{2D_y}{\lambda} \left(1-e^{-Qt} \right)  + 2QD_y \mathcal{X}(t,Q,\lambda,0)
\end{multline}
\begin{multline}
\langle x^2(t) \rangle _{\lambda =0} = x_0^2 e^{-Qt} + y_0^2 \left(1-e^{-Qt} \right)+D_y t \\
+ \frac{2D_x-D_y}{Q} \left(1-e^{-Qt} \right) 
\end{multline}
\begin{multline}
\langle x^2(t) \rangle _{\lambda >0} = x_0^2 e^{-Qt} + y_0^2 \left(1-e^{-Qt} \right) + \frac{2D_x}{Q} \left(1-e^{-Qt} \right)  \\
+ \frac{2D_y}{\lambda} \left(1-e^{-Qt} \right)  - 2QD_y  \mathcal{X}(t,Q,\lambda,1)
\end{multline}
where we define the function $\mathcal{X}(t,Q,\lambda,n_0)$ as,
\begin{multline}
\mathcal{X}(t,Q,\lambda, n_0) = \sum_{n=n_0}^{\infty} \frac{e^{-Qt}-e^{-|\lambda|n t} \left(1-t \left(Q- |\lambda| n \right) \right)}{ \left( Q- | \lambda | n \right)^2} \\
\times \left(1 - \mathbf{1}_{Q,n |\lambda|} \right) + \frac{t^2 e^{-Qt}}{2} \mathbf{1}_{Q,n |\lambda |} \; .
\end{multline}

These expressions were compared with simulations and are shown in Fig. 2 of the main text.

%


\section{Mean First Passage Time}
In this section we show the method used to compute the first passage time. Let us denote as $z=x-y$ the relative distance between the prey and the predator. Notice that the encounter between the prey and the predator corresponds to $z=0$. First, let us consider the situation in which $Q=0$. It is widely known that in this case the survival probability is \cite{redner2001guide},
\begin{equation}
\mathcal{S}(z_0, t)=\mathbf{Erf} \left(\frac{z_0}{\sqrt{4D t}} \right) \; . \label{eq:S0}
\end{equation}
Here $z_0=x_0-y_0$ is the initial position and $D=D_x+D_y$. Now, let us assume $Q\neq0$ and that only one reset occurs to $z=z_1$. Then, the survival probability of the whole process is, inf fact, the convolution of the subprocess before the reset multiplied by the probability the reset does not occur and the subprocess after the reset multiplied, again, by the probability the reset does not occur. Although we are considering only one reset, we include the probability the reset does not occur in both terms of the convolution because in the end we will consider the limit for infinite number of resets. Therefore, the previous ideas yield the following:
\begin{eqnarray}
\mathcal{S}^{(1)}(z_0,z_1, t)&=&Q\int_0^t dt_1 e^{-Qt_1} \mathcal{S}(z_1,t_1) \nonumber \\
 & \times & \int_0^t dt_0 e^{-Qt_0} \mathcal{S}(z_0, t_0)\delta(t-t_1-t_0) \; .
\end{eqnarray}
Generalizing to $n$ resets is straightforward. Hence, for $n$ resets we may write,
\begin{multline}
\mathcal{S}^{(n)}(z_0,\lbrace z_i \rbrace,t)=\prod_{i=1}^n \int_0^t dt_i e^{-Qt_i} Q \mathcal{S}(z_i,t_i)  \\
\times  \int_0^t  dt_0 e^{-Qt_0} \mathcal{S}(z_0,t_0) \delta \left(t-\sum_{i=0}^n t_i \right) \; . \label{eq:Sn}
\end{multline}
Now, applying Laplace transform, which we denote with a $\widetilde{\bullet}$, on the previous Eq. \eqref{eq:Sn} yields
\begin{equation}
\widetilde{\mathcal{S}}^{(n)}(z_0,\lbrace z_i \rbrace,u)=\prod_{i=1}^n \left[ Q \mathcal{S}(z_i,u+Q) \right] \mathcal{S}(z_0,Q+u) \; .
\end{equation}
The survival probability given any number of resets, $S_Q^{(n)}$, yields
\begin{multline}
\widetilde{\mathcal{S}}_Q(z_0  ,\lbrace z_i \rbrace  ,u )=\widetilde{\mathcal{S}}(z_0,Q+u)+\sum_{n=1}^{\infty} \widetilde{\mathcal{S}}^{(n)}(z_0,u)  \\
= \widetilde{\mathcal{S}}(z_0,Q+u)\left(1+\sum_{n=1}^{\infty} \prod_{i=1}^n \left[ Q \widetilde{\mathcal{S}}(z_i,u+Q) \right] \right) \; . \label{eq:S}
\end{multline}
From the Laplace transform of the survival probability one may obtain the mean first passage time by taking the limit $u \rightarrow 0$.
Moreover, the Laplace transform of the survival probability given that there is no reset (see Eq. \eqref{eq:S0}) yields,
\begin{equation}
\widetilde{\mathcal{S}}_Q(z_0,u)=\frac{1}{u}\left(1-\exp\left(-\sqrt{\frac{u}{D}} \vert z_0 \vert \right) \right) \; .
\end{equation}
Thus, Eq. \eqref{eq:S} becomes,
\begin{multline}
\widetilde{\mathcal{S}}_Q(z_0,\lbrace z_i \rbrace,u)=\frac{1}{Q+u}\left(1-e^{-\sqrt{\frac{u+Q}{D}} \vert z_0 \vert } \right) \\ \left(1+\sum_{n=1}^{\infty} \prod_{i=1}^n \left[  \frac{Q}{Q+u}\left(1-e^{-\sqrt{\frac{u+Q}{D}} \vert z_i \vert } \right) \right] \right) \; . \label{eq:S2}
\end{multline}
Let us now consider some well-known scenarios to support the usefulness of Eq. \eqref{eq:S2}. First, notice that Eq. \eqref{eq:S2} diverges in the limit when $Q \rightarrow 0$ and $u \rightarrow 0$, in agreement with the mean first passage time when there are no resets. Now, recall that $z_i$ is the relative distance once the $i^{th}$ reset takes place. Then, for the case where $\lambda \rightarrow -\infty$, i.e., the predator resets to the present position of the prey, we have $z_i=0$ for all $i$. Thus, Eq. \eqref{eq:S2} becomes,
\begin{equation}
\widetilde{\mathcal{S}}_Q(z_0,\lbrace 0 \rbrace,u)=\frac{1}{Q+u}\left(1-e^{-\sqrt{\frac{u+Q}{D}} \vert z_0 \vert } \right)  \; . \label{eq:Sex1}
\end{equation}
The mean first passage time yields, by taking the limit $u \rightarrow 0$,
\begin{equation}
\langle t \rangle_{\lambda \rightarrow -\infty}= \frac{1}{Q}\left(1-e^{-\sqrt{\frac{Q}{D}} \vert z_0 \vert } \right) \; .
\end{equation}
which may also be obtained by a backward-Fokker-Planck procedure. Additionally, the previous Eq. has been plotted in Fig. 3 of the main text.

Another interesting case corresponds to when $z_i=z_0$ for all $i$ which means that the relative distance is always the same after each reset, then Eq. \eqref{eq:S2} becomes
\begin{multline}
\widetilde{\mathcal{S}}_Q(z_0,\lbrace z_i \rbrace,u)=\frac{1}{Q+u}\left(1-e^{-\sqrt{\frac{u+Q}{D}} \vert z_0 \vert } \right) \\ \left(1+\sum_{n=1}^{\infty}  \left[ \frac{Q}{Q+u}\left(1-e^{-\sqrt{\frac{u+Q}{D}} \vert z_i \vert } \right) \right]^n \right) \; .
\end{multline}
The summation in the previous Eq. is a geometric summation and term inside is between $0$ and $1$, hence, the summation converges. It is possible to show that taking the limit $u \rightarrow 0$ yields the mean first passage time of a Brownian particle hitting the origin given that it resets to the same position with rate $Q$, namely,
\begin{equation}
\langle t \rangle_{\forall_{i} z_i  = z_0}=\frac{1}{Q}\left(e^{\sqrt{\frac{Q}{D}} \vert z_0 \vert } -1 \right) \; ,
\end{equation}

In perfect agreement with Ref. \cite{evans2011diffusion}. Now, the actual value of $S(z_0, \lbrace z_i \rbrace, u)$ and, thus, the mean first passage time will depend on the 
actual values of $z_i$ which are random variables. Hence, the exact result would be rather cumbersome to obtain. We, however, circumvent this issue by truncating the summation 
up to the mean number of resets for the first encounter. In the following section we discuss this.

\section{Mean number of resets}
In this section we show the derivation to obtain the mean number of resets for the first encounter for $\lambda \rightarrow \pm \infty $. Let us first consider the case $\lambda \rightarrow - \infty$, i.e., the predator relocates to the present position of the prey when the reset occurs. Notice that the mean time between resets is simply $1/Q$ and, henceforth, we consider this approximation in our derivations. Therefore, in a coarse-grained approximation, we may think of the probability of first encounter before the first (and only) reset occurs, which we denote as $p(Q)$, given by,
\begin{equation}
p(Q)=\int_0^{1/Q} \frac{|z_0|}{\sqrt{4\pi D t^3}} e^{-\frac{z^2}{4Dt}} = \mathbf{Erfc} \left(\frac{|z_0|}{2} \sqrt{\frac{Q}{D}} \right) \; . \label{eq:19}
\end{equation} 
Conversely, the probability of first encounter after the first reset would be the probability the encounter did not happen before the first encounter, $1-p(Q)$, multiplied by the probability the encounter occurred after the first reset, which is equal to $1$ given that $\lambda \rightarrow -\infty$. It is trivial to show that the probability for the encounter happening before or after the first reset is equal to $1$. Then, the mean number of reset for the first encounter, $ \langle n(Q,\lambda \rightarrow -\infty) \rangle$, is given by,
\begin{equation}
\langle n(Q,\lambda \rightarrow -\infty) \rangle = 1-p(Q)   \; . \label{eq:meanResMInf}
\end{equation}
In Fig. 4 of the main text we compare Eq. \eqref{eq:meanResMInf} with the mean number of resets before the first encounter obtained from our simulations.

\begin{figure}[hbtp]
\includegraphics[width=\columnwidth]{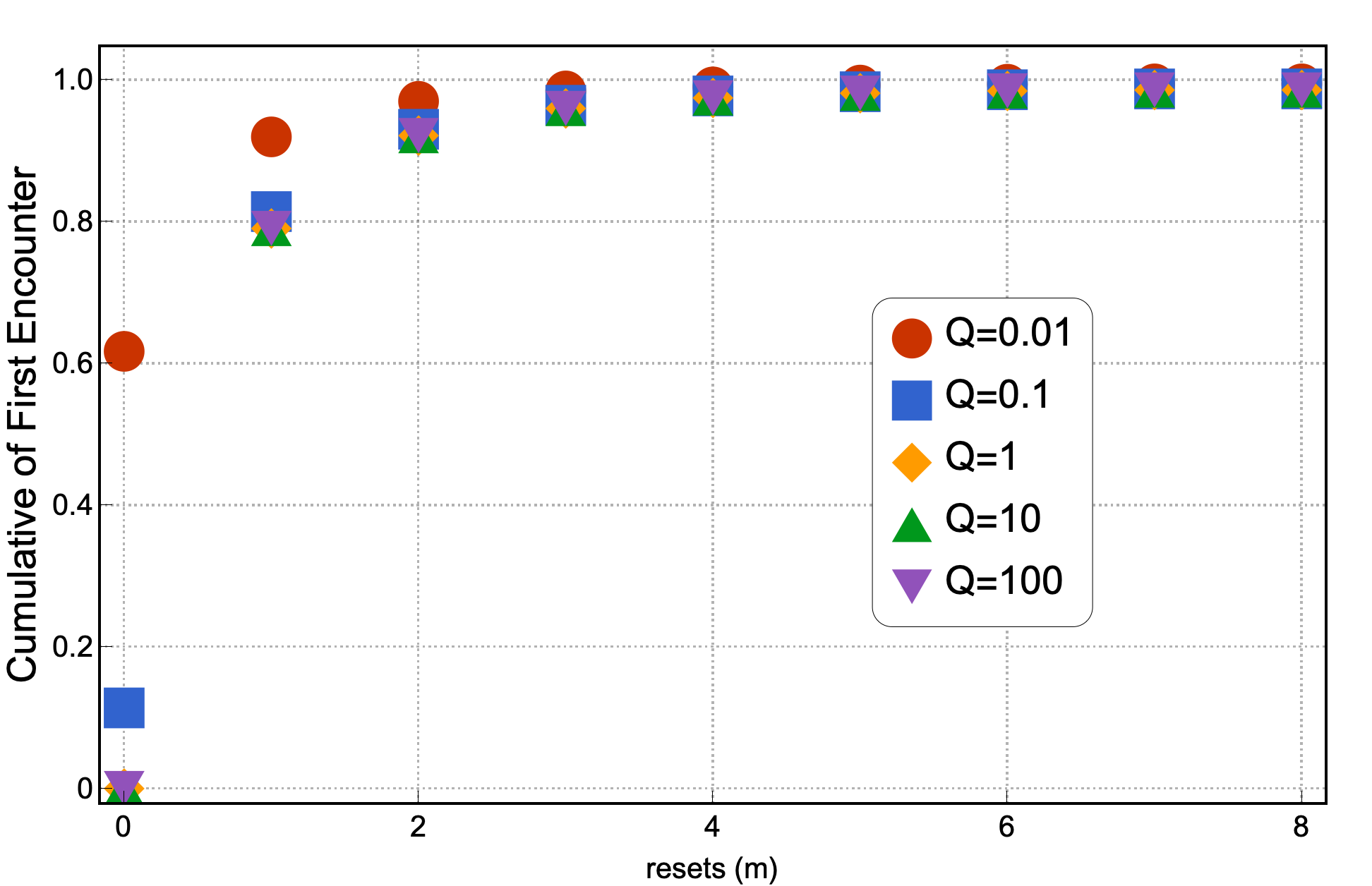}
\caption{Cumulative of the first encounter probability \textit{vs} the number of resets, for different values of $Q$ (see legend). Notice that for $Q\geq 0.1$ the curves collapse at $m=2$, which is right before the cumulative saturates. On the other hand, for $Q=0.01$, the cumulative saturates before. This implies that after $3$ resets, the first encounter probability is essentially $1$. This explains why the mean number of resets for the first encounter reaches a plateau as shown in Fig. 4 of the main text and, hence, supports the approximation of truncating the infinite sum in Eq. (11) of the main text. This result was obtained using the saddle point method (see text for further details). } \label{Fig:CumvsRes}
\end{figure}

Let us now turn our attention to the case where $\lambda \rightarrow \infty$, i.e., the predator always resets to the initial position of the prey. The pathway is the same as in the previous case. In order to simplify the derivation, it is helpful to consider a large but finit number of resets. 
Then, the first encounter probability after the first reset and before the second reset is given by,
\begin{multline}
P(n=1)= \mathbf{Erf} \left(\frac{|z_0|}{2} \sqrt{\frac{Q}{D}} \right) \mathbf{Erfc} \left(\frac{|z_1|}{2} \sqrt{\frac{Q}{D}} \right) \\
\times A \left(n, z_2,...,z_N \right) \; . \label{eq:P1}
\end{multline}
The first term in Eq. \eqref{eq:P1} corresponds to the non-encounter probability before the first reset whereas the second term corresponds to the encounter probability between the first and second reset. The last term corresponds to the probability of encounter after the second reset and takes into account all possible combination. In the case where $z_i=z_0$ for all $i>0$ it is trivial to show that the last term is $\left(p(Q)+(1-p(Q)) \right)^N =1$. In the present case, the exact form of $A(n,\lbrace z_i \rbrace)$ takes into account all the encounter probabilities after the second reset and is rather complex since it depends on the actual values of $\lbrace z_i \rbrace$. In any case, by normalization the value must be equal to $1$. Then, the probability of first encounter after the $m$th reset and before the $m$th$+1$ is given by,
\begin{multline}
P(m)=\mathbf{Erf} \left(\frac{|z_0|}{2} \sqrt{\frac{Q}{D}} \right) \prod_{i=1}^{m-1} \left( \mathbf{Erf} \left(\frac{|z_i|}{2} \sqrt{\frac{Q}{D}} \right) \right) \\ 
\times \mathbf{Erfc} \left(\frac{|z_m|}{2} \sqrt{\frac{Q}{D}} \right) \; . \label{eq:Pm}
\end{multline}
The mean number of resets for the first encounter, $ \langle n(Q,\lambda \rightarrow \infty) \rangle$, is simply,
\begin{equation}
 \langle n(Q,\lambda \rightarrow \infty) \rangle = \sum_{m=0}^{\infty} P(m) m \; . \label{eq:meanResInf}
\end{equation}
We may further ask \textit{How many resets are required for the first encounter to happen?} To this end, we compute the cumulative of Eq. \eqref{eq:Pm}. The fact of the matter is that the $z_i$ are random variables, thus, in principle, one needs to average over these variables. To circumvent this issue, we consider $z_i$ to be Gaussian random variables with mean squared displacement equal to $\sqrt{2D_y i/Q}$ and then we use the saddle point method. Therefore, we obtain,
\begin{multline}
P_{sp}(m)=\mathbf{Erf} \left(\frac{|z_0|}{2} \sqrt{\frac{Q}{D}} \right) \\
\times \prod_{i=1}^{m-1} \left( \sqrt{\frac{Q}{2D_y \frac{d^2 B_i(z)}{d z^2}\vline _{z=z^*}}} e^{-B_i(z^*)} \right) \\
\times \left(1-\sqrt{\frac{Q}{2D_y \frac{d^2 B_m(z)}{d z^2}\vline _{z=z^*}}} e^{-B_m(z^*)}  \right) \; , \label{eq:SP}
\end{multline}
where
\begin{equation}
B_i(z)=\frac{z^2 Q}{4D_y i} - \log \left( \mathbf{Erf} \left(\frac{|z|}{2} \sqrt{\frac{Q}{D}} \right) \right) \; .
\end{equation}
We denote as $z=z^{*}$ the value for which the derivative of $B_i(z)$ with respecto to $z$ equates to zero. Using Eq. \eqref{eq:SP} we construct the cumulative of first encounter, i.e., $\sum P_{sp}(m)$ and we plot this in Fig. \ref{Fig:CumvsRes} to answer the aforementioned question. Notice that after two resets, the cumulative saturates regardless of the value of $Q$. 

Using Eq. \eqref{eq:meanResInf} together with Eqs. \eqref{eq:P1} and \eqref{eq:Pm} we are able to give an approximation for the mean number of resets for the first encounter. Additionally, we made the approximation $z_i=\sqrt{2D_y i/Q}$ and we compared it with our simulations in Fig. 4 of the main text for a wide range of values of $Q$ and the agreement is very good.

\section{Numerical simulations}
In this section we describe the implementation our simulations. We first discretized the Langevin Eqs. shown in the main text. In the case where no capture was considered we ran the simulations for different $Q$ values and negative values of $\lambda$. These results were then compared with Eqs. \eqref{eq:LambdaNeg} and \eqref{eq:Lambda0} in Fig. 2 of the main text and shows excellent agreement. We then used this simulation for the statistics of the first encounter time. However, in order to correctly sample between resets, the time step of the simulation has to be much smaller the the characteristic reset time, $1/Q$, to properly sample between resets. This then leads to a rather dynamical time step and, additionally, as the time step decreases the computational time increases. Thus, going in this direction implies a trade-off between correct sampling and computational time. For these reasons, we instead performed a Kinetic Monte Carlo type simulation which reduces the computational time significantly and is fairly easy to implement for the cases $\lambda \rightarrow \pm \infty$.

First, we generate a first passage time random variable, $t_0$, given the initial relative distance $z_0$,  between the predator and the prey, from a L\'evy-Smirnov distribution as well as a reset time random 
variable $\tau_0$, given the fixed reset rate $Q$, from a Poisson distribution. Then, if $t_0>\tau_0$, we keep repeating the previous step and denote the new first passage random 
variable as $t_n$ obtained from a L\'evy-Smirnov distribution, given a new relative distance right after the reset, $z_n$, and the new reset time random variable as $\tau_n$, where $n$ 
denotes the number of reset events that have occurred since the process started. If $t_n< \tau_n$, then the process stops and the first passage time random variable for that sample is 
$\sum_{i=0}^{n-1} \tau_i+t_{n}$. In general, the $z_n$'s are also random variables with unknown distributions. But, in the case where $\lambda \rightarrow -\infty$, $z_n=0$ for all $n>0$, conversely when $\lambda \rightarrow \infty $, then $z_n$ is a Gaussian random variable with mean equal to zero and mean squared displacement equal to $2D_y \sum_{i=0}^{n-1} \tau_n$. The results were then compared with our analytical approximation in Fig. 3 in the main text.


\end{document}